\documentclass[a4paper,11pt]{article}
\pdfoutput=1 

\usepackage{jinstpub} 

\usepackage{lineno}
\usepackage{epsfig}
\usepackage{epstopdf}
\usepackage{color} 
\usepackage{url}

\newcommand{\xspace}{~}
\newcommand{\onbb}{$0\nu2\beta$\xspace}

\newcommand{\LMOenr}{Li$_{2}${}$^{100}$MoO$_4$\xspace}

\title{A CUPID \LMOenr scintillating bolometer tested in the CROSS underground facility}

\author[a]{A.~Armatol,}
\author[a]{E.~Armengaud,}
\author[b]{W.~Armstrong,}
\author[c]{C.~Augier,}
\author[d]{F.~T.~Avignone~III,}
\author[e]{O.~Azzolini,}
\author[f]{I.~C.~Bandac,}
\author[g]{A.~S.~Barabash,}
\author[h]{G.~Bari,}
\author[i,j]{A.~Barresi,}
% ...

\affiliation[a]{IRFU, CEA, Universit\'e Paris-Saclay, F-91191 Gif-sur-Yvette, France}
\affiliation[b]{Argonne National Laboratory, Argonne, IL 60439, USA}
\affiliation[c]{Univ Lyon, Universit\'{e} Lyon 1, CNRS/IN2P3, IP2I-Lyon, F-69622, Villeurbanne, France}
\affiliation[d]{Department of Physics and Astronomy, University of South Carolina, Columbia, SC 29208, USA}
\affiliation[e]{INFN -- Laboratori Nazionali di Legnaro, Legnaro (Padova) I-35020, Italy}
\affiliation[f]{Laboratorio Subterr\'aneo de Canfranc, 22880 Canfranc-Estaci\'on, Spain}
\affiliation[g]{National Research Centre Kurchatov Institute, Institute of Theoretical and Experimental Physics, 117218 Moscow, Russia}
\affiliation[h]{INFN -- Sezione di Bologna, Bologna I-40127, Italy}
\affiliation[i]{INFN -- Sezione di Milano Bicocca, Milano I-20126, Italy}
\affiliation[j]{Dipartimento di Fisica, Universit\`{a} di Milano-Bicocca, Milano I-20126, Italy}
% ...

\emailAdd{cupid.publications@lngs.infn.it}
\emailAdd{cross.publications@lsc-canfranc.es}

\abstract{A scintillating bolometer based on a large cubic \LMOenr crystal (45 mm side) and a Ge wafer (scintillation detector) has been operated in the CROSS cryogenic facility at the Canfranc underground laboratory in Spain. The dual-readout detector is a prototype of the technology that will be used in the next-generation \onbb experiment CUPID. The measurements were performed at 18 and 12 mK temperature in a pulse tube dilution refrigerator. This setup utilizes the same technology as the CUORE cryostat that will host CUPID and so represents an accurate estimation of the expected performance. 
The Li$_{2}${}$^{100}$MoO$_4$ bolometer shows a high energy resolution of 6 keV FWHM at the 2615 keV $\gamma$ line. The detection of scintillation light for each event triggered by the \LMOenr bolometer allowed for a full separation ($\sim$8$\sigma$) between $\gamma$($\beta$) and $\alpha$ events above 2 MeV. The \LMOenr crystal also shows a high internal radiopurity with $^{228}$Th and $^{226}$Ra activities of less than 3 and 8 $\mu$Bq/kg, respectively. 
Taking also into account the advantage of a more compact and massive detector array, which can be  made of cubic-shaped crystals (compared to the cylindrical ones), this test demonstrates the great potential of cubic \LMOenr scintillating bolometers for high-sensitivity searches for the $^{100}$Mo \onbb decay in CROSS and CUPID projects.}

\keywords{Double-beta decay, Cryogenic detector, Bolometer, Crystal scintillator, Lithium molybdate, Particle identification,  Radiopurity}

\arxivnumber{} 

\begin{document}
\maketitle
\flushbottom

\section{Introduction}
\label{sec:intro}

Neutrinoless double-beta ($0\nu2\beta$) decay is a unique probe of new physics beyond the Standard Model \cite{Vergados:2016,Dolinski:2019} and the observation of this process, suggested about 80 years ago but not yet detected (in contrast to two-neutrino double-beta ($2\nu2\beta$) decay \cite{Barabash:2020}), would conclusively demonstrate lepton number violation and the Majorana nature of neutrinos (i.e. a particle that is equal to its own anti-particle). 

The bolometric technology, which relies on the use of low-temperature calorimeters acting simultaneously as a $2\beta$ source and a detector, is among the few experimental approaches providing world-leading sensitivity to $0\nu2\beta$ decay to-date \cite{Dolinski:2019}. In addition to high detection efficiency of the ``$2\beta$ source = detector'' technique, bolometers offer high energy resolution, scalability to a large detector mass via arrays of modules, and the possibility to use different and radiopure materials containing the most promising $2\beta$ isotopes (e.g. see \cite{Pirro:2017,Poda:2017a,Bellini:2018}). Additionally, recent technological advances have demonstrated the ability to do particle identification \cite{Pirro:2017,Poda:2017a,Bellini:2018}, allowing for a reduction of backgrounds in the signal region of interest by multiple orders of magnitude. 

Bolometric techniques for $0\nu2\beta$ decay searches have been developed for about 30 years and have resulted in the first tonne-scale bolometric experiment CUORE (Cryogenic Underground Observatory of Rare Events) \cite{Adams:2020}. CUORE has been in operation at the Gran Sasso underground laboratory (Italy) since 2017, searching for $0\nu2\beta$ decay of $^{130}$Te ($Q$-value of the $2\beta$ transition, $Q_{2\beta}$, is 2528~keV \cite{Wang:2017a}). In spite of this extraordinary achievement, the CUORE $0\nu2\beta$ sensitivity is limited by a background ($\sim$10$^{-2}$ counts/yr/kg/keV) coming from alpha decays at surfaces despite the highly radiopure materials used for the detector construction. This is due to the use of pure thermal detectors based on tellurium dioxide crystals (TeO$_2$; 34\% of $^{130}$Te in natural tellurium \cite{Meija:2016}) which have the same bolometric response irrespective of the type of the particle interaction \cite{Bellini:2010}. 

CUPID (CUORE Upgrade with Particle IDentification) is a proposed next-generation $0\nu2\beta$ bolometric experiment \cite{Wang:2015raa}, which will reuse the CUORE infrastructure for the operation of a similar-scale isotopically enriched detector with a background $\sim$10$^{-4}$ counts/yr/kg/keV in the region of interest, thus probing  $0\nu2\beta$ decay in so-called ``zero-background'' conditions. 
The suppression of the alpha-induced background to a negligible level (i.e. 99.9\% of alpha events rejection), while keeping almost 100\% of the signal efficiency, is required for particle identification technology. 
The detector performance is expected to be similar to CUORE and predecessors, with a 5 keV FWHM at $Q_{2\beta}$ as a goal.
The activities of $^{226}$Ra, $^{228}$Th, and $^{232}$Th in the enriched bolometers are required to be less than 10 $\mu$Bq/kg, making the contribution of the U/Th crystal bulk activity to be below $\sim$10$^{-4}$~counts/yr/kg/keV. The total bulk radioactivity of the crystals should not exceed the mBq/kg level to avoid impacting the detector operation and background with pile-ups \cite{Chernyak:2012,Chernyak:2014,Chernyak:2017,Poda:2017a}.

Four isotopes, $^{82}$Se ($Q_{2\beta}$ = 2998 keV \cite{Wang:2017a}), $^{100}$Mo (3034 keV \cite{Wang:2017a}), $^{116}$Cd (2813 keV \cite{Wang:2017a}) and $^{130}$Te, were considered in the CUPID R\&D program \cite{Wang:2015taa} as isotopes of interest to be embedded in the CUPID detector for the following reasons: 
\begin{itemize}
\item  The $0\nu2\beta$ decay energy of these isotopes (except for $^{130}$Te) is greater than 2.6 MeV, the end-point of the most energetic intense natural $\gamma$-ray radiation; 
\item  Enrichment is available at a large amount and reasonable cost; 
\item  Compounds containing these isotopes can be grown into single crystals usable for cryogenic applications; 
\item  Some of Se-, Mo-, or Cd-containing crystals are also reasonably efficient low-temperature scintillators. The detection of scintillation light using an auxiliary optical bolometer in coincidences with the measurement of particle-induced energy release in the scintillating absorber is a viable tool for particle identification. This technique can also be applied for poorly or non-scintillating crystals, as TeO$_2$, to detect Cherenkov radiation allowing particle identification (however, more performing light detectors are demanded to detect a tiny signal).  \end{itemize}
Efficient alpha background rejection has been demonstrated with detectors containing each of these isotopes \cite{Wang:2015taa,Poda:2017,Azzolini:2019tta,Armengaud:2017,Helis:2020}. This paves the way for a future study of \onbb across multiple isotopes \cite{Giuliani:2018} in case a discovery is made. Based on performance and cost, CUPID selected $^{100}$Mo embedded in lithium molybdate (Li$_2$MoO$_4$) scintillating crystals \cite{CUPIDInterestGroup:2019inu}. 

The technology of $^{100}$Mo-enriched lithium molybdate (Li$_{2}${}$^{100}$MoO$_4$) scintillating bolometers has been recently developed within the LUMINEU project and it provides \cite{Armengaud:2017,Grigorieva:2017}: \begin{itemize}
\item A know-how for the mass production of high-quality large radiopure crystals with only few \% losses of the enriched material; 
\item The fabrication of a detector module (which can be easily put into array) with energy resolution comparable to that of TeO$_2$ bolometers, but with a significantly higher $\alpha$ rejection efficiency (e.g. see in \cite{Poda:2017a}). 
\end{itemize}
Excellent performances of \LMOenr scintillating bolometers based on cylindrical crystals ($\oslash$44$\times$45 mm, $\sim$0.21 kg, $\sim$97\% enrichment in $^{100}$Mo) have been demonstrated in single-module and 4-crystal-array tests of LUMINEU  \cite{Armengaud:2017,Poda:2017a,Armengaud:2020b} at the Gran Sasso and Modane (LSM; France) underground laboratories. These results have been recently confirmed by the CUPID-Mo experiment \cite{Armengaud:2020a,Schmidt:2020,Poda:2020} on the scale of a 20-detector array operated at the LSM.
Furthermore, the crystal production protocol, adopted by LUMINEU and CUPID-Mo, has been used for the fabrication of 32 \LMOenr crystals 0.28 kg each (the average enrichment in $^{100}$Mo is 97.7(3)\%) for the CROSS (Cryogenic Rare-event Observatory with Surface Sensitivity) $0\nu2\beta$ experiment \cite{Bandac:2020}.
 
CROSS, considered as a part of CUPID R\&D, is a project aiming at the development of \LMOenr  
and $^{130}$TeO$_2$ surface-coated bolometers capable of identifying a near surface particle interaction via pulse-shape analysis \cite{Bandac:2020}. A key ingredient of the CROSS technology is crystal-surface coating with a superconducting material to modify the signal pulse-shape for an event occurring at its proximity. The CROSS detector performance and radiopurity should be in compliance with CUPID requirements. The feasibility of a highly-efficient identification of near-surface $\alpha$ interactions has recently been demonstrated in multiple tests of the CROSS prototypes \cite{Bandac:2020,Khalife:2020,Khalife:2020a,Zolotarova:2020a}. A final validation of the technology is planned to be realized as a $0\nu2\beta$ experiment with at least 32 \LMOenr bolometers (the addition of the 20 crystals from CUPID-Mo are now also in consideration), hosted in a dedicated cryostat at the Canfranc underground laboratory (Spain). The sensitivity of this medium-scale demonstrator \cite{Bandac:2020} is expected to be on the level of the leading $0\nu2\beta$ experiments, which have masses larger by a factor 10--100.

In contrast to LUMINEU and CUPID-Mo, CROSS is going to use cubic \LMOenr elements with a 45 mm side. The choice of a cubic shape is driven by the possibility to realize a more compact array structure, which allows to deploy a $\sim30\%$ higher isotope mass in the available experimental volume, and yields an enhanced efficiency in rejecting background-like events that release energy in neighboring crystals (coincidences). Indeed, a volume (i.e. mass) of a cylindrical crystal with a diameter and height equal to the side of the cubic one is almost 30\% less (similar to CUPID-Mo vs. CROSS crystals). It is also evident that the efficiency of coincidences between larger, particularly neighbor, detectors would be increased too. 
The CROSS development of an array of cubic \LMOenr bolometers is also an important benchmark for the design of the final CUPID structure, initially considered to be based on cylindrical ($\oslash50\times50$~mm) \LMOenr scintillating bolometers \cite{CUPIDInterestGroup:2019inu}. 
Before using the crystals in the CROSS and CUPID \onbb experiments, it is necessary to perform low-temperature test(s) to demonstrate that: 
\begin{itemize}
\item Bolometric and spectrometric performances of cubic-shaped \LMOenr scintillating bolometers, operated in modern pulse-tube cryostats with possible vibration disturbances, are similar to those of cylindrical \LMOenr detectors tested in dry and/or wet dilution refrigerators; 
\item Scintillation light yield of the cubic-shaped and cylindrical crystals is similar, thus providing a highly efficient particle identification;
\item Radioactive contamination of cubic \LMOenr crystals is compatible to that of cylindrical ones. 
\end{itemize}
With these goals in mind, we realized a first investigation of a scintillating bolometer based on a large-volume ($\sim$90~cm$^3$) cubic-shaped \LMOenr crystal, described in the present paper. This study is undertaken as part of both the CROSS and CUPID R\&D programs.

\section{Detector construction and operation}
\label{sec:experiment} 

\subsection{Li$_{2}${}$^{100}$MoO$_4$ scintillating bolometer fabrication}

\begin{figure}[hbt]
\centering
\resizebox{0.55\textwidth}{!}{\includegraphics{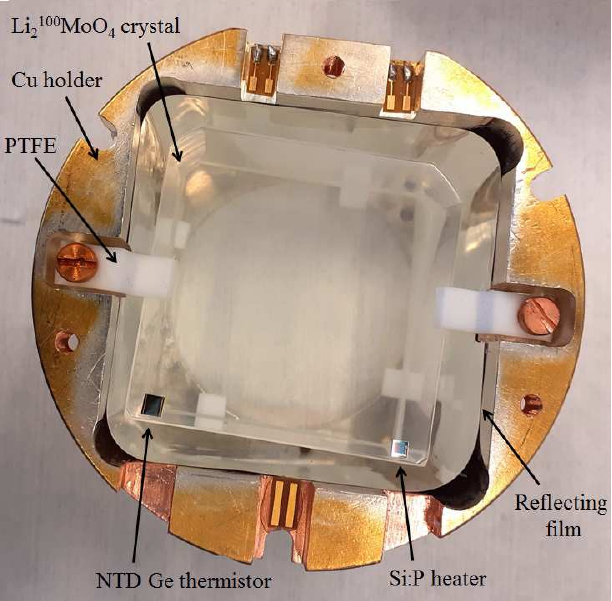}}
\caption{A photograph of the partially assembled \LMOenr scintillating bolometer; the construction elements (see text) are labeled. A $\oslash45$ mm hole at the bottom of the holder, visible in transparency, acts as an entrance window for the \LMOenr scintillation light to be registered by a bolometric Ge light detector.}
\label{fig:Detector}
\end{figure}

We construct a dual-readout cryogenic particle detector from a primary scintillating absorber and a light detector. The \LMOenr absorber consists of a 45$\times$45$\times$45~mm crystal ($\sim$98\% enrichment in $^{100}$Mo) of mass 279.42 g. 
We randomly chose the sample from the batch of 32 identical crystals. The crystals were grown starting from purified $^{100}$Mo powder and using the low-temperature-gradient technique at the Nikolaev Institute of Inorganic Chemistry (Novosibirsk, Russia). 
The \LMOenr samples are not perfectly cubic-shaped\footnote{There are edge chamfers on the crystals, see Fig.~\ref{fig:Detector}.} due to not optimal growing conditions (in particular, the platinum crucible size was not large enough), adapted for the growth of up to $\oslash$50~mm crystal boules \cite{Grigorieva:2017,Armengaud:2017,Armengaud:2020a}. 
A Neutron Transmutation Doped (NTD) Ge thermistor \cite{Haller:1994} with a size of 3$\times$3$\times$1~mm and a P-doped Si chip \cite{Andreotti:2012} were epoxy-glued on the crystal top. The dependency of the NTD Ge resistance on temperature can be approximated as $R(T) = R_0 \cdot e^{(T_0/T )^{0.5}}$ with the parameters $T_0$ $\sim$ 3.8~K and $R_0$ $\sim$ 1.5~$\Omega$. 
The Si chip is used as a resistive element to periodically inject constant energy pulses used for off-line stabilization of the bolometric response \cite{Alessandrello:1998}. 
The crystal holder is made from copper to host cubic crystals of up to 5 cm side and optical bolometers at their top and/or bottom \cite{Novati:2018,Berge:2018,Zolotarova:2020}. As for the light detector (LD) we use a SiO-coated Ge wafer of 44 mm diameter and 0.175 mm thickness, instrumented with a 3$\times$1$\times$1~mm NTD. The LD was mounted in the copper holder near the \LMOenr detector (LMO).

The crystal is fixed inside the copper holder by PTFE (polytetrafluoroethylene) pieces, as seen in Fig.~\ref{fig:Detector}. The PTFE supports act also as thermal contacts to the heat sink of the cryostat. The Cu holder of the detector is coated with Au to avoid oxidation, while the internal part of the holder is also coated with Ag to improve light reflection. 
The Li$_2${}$^{100}$MoO$_4$ crystal inside the Cu housing is surrounded with a Vikuiti{\texttrademark} reflecting film, which is the same used in LUMINEU and CUPID-Mo. The NTD Ge is wire-bonded with Au wires, while the heater is bonded with Al wires. 
A $^{238}$U/$^{234}$U source is placed on the holder's top cap. The source was obtained by depositing an uranium-containing liquid drop on a thin copper substrate that was then dried by evaporation. Part of the alpha particles (as well as nuclear recoils) emitted are degraded in energy. The LD is fabricated in the same way using a dedicated Cu holder, three PTFE elements, and an NTD Ge sensor glued. A $^{55}$Fe X-ray source is placed close to the LD to irradiate the Ge surface opposite to the \LMOenr absorber.

\subsection{Low-temperature underground measurements}

We tested the detector in the CROSS Cryogenic Underground (C2U) facility \cite{Olivieri:2020}, in operation at the Canfranc laboratory (Spain) since April 2019. The cryostat is placed inside a Faraday cage with acoustic isolation, formerly used by the ROSEBUD dark matter experiment \cite{Cebrian:2004}. The set-up operates a pulse-tube (Cryomech PT415) based dilution refrigerator, developed by CryoConcept (France), which is also assisted by Ultra Quiet Technology{\texttrademark} (UQT) to mitigate vibrations \cite{Olivieri:2017}. 
During the cryostat commissioning, it was found the UQT to efficiently reduce vibrations in the vertical direction, but not as much horizontally \cite{Olivieri:2020} resulting in a noise excess affecting the bolometric performance \cite{Helis:2020}. Thus, the hybrid bolometer was spring-suspended from the detector plate.
In order to reduce the environmental background, the cryostat is surrounded externally by a 25 cm thick low-radioactivity lead shield. Moreover, the detector volume inside the cryostat is shielded from the dilution unit and cryostat upper parts with a 13 cm thick disk made of sandwiched lead and copper (120 kg total mass). The shielding of the set-up has not been completed yet, in particular an anti-radon Plexiglas box (to be flushed with a deradonized air) and a muon veto will be installed soon.

The signal readout is based on a low-noise room-temperature DC front-end electronics \cite{Arnaboldi:2002} tracing back to the Cuoricino experiment. The data acquisition (DAQ) is a new design candidate for CUPID \cite{Carniti:2020} and consists of two 12-channel boards with a programmable 6-pole Bessel-Thomson anti-aliasing filter and integrated 24-bit ADC. A cut-off frequency of the low-pass filter can be set from 24 Hz up to 2.5 kHz. With the 24 bit ADC resolution, the input noise is not limited by the ADC even with the lowest gain value set at a programmable-gain amplifier (PGA). An additional advantage of such ADC resolution is that the PGA stage can be made much simpler or removed, with less power consumption, cost, and space. The sampling rate up to 25 kS/s can be set (250~kS/s with half of channels). The ADC-digitized continuous data are readout by an external FPGA (field-programmable gate array) module and then transferred to a personal computer via Ethernet. The DAQ control is done with the help of a MATLAB-based graphical user interface program. The monitoring on-line of the data quality is realized as a LabVIEW application.

We collected data from the end of December 2019 until the beginning of April 2020. 
The measurements were performed at temperatures 18 and 12 mK. We periodically calibrated the LMO by inserting a thoriated tungsten wire inside the lead shield. We chose the working points for both operational temperatures to be a few nA current on the NTD sensor resulting in a few M$\Omega$ resistance. The data are sampled continuously at a 2 kS/s sampling rate, and the full data stream is written to disk for offline analysis. The Bessel-Thomson cut-off frequency was set at 300 Hz, as a compromise between the bandwidth of the LMO (slow) and LD (relatively fast).  

We used around-3-week-long stable periods of data for the analysis at each regulated temperature, not affected by external events (e.g. power cuts). We select 314 h of physics data at each  temperature and 65 and 220 h of the $^{232}$Th calibration data at 18 and 12 mK, respectively. 

The acquired data are triggered offline to tag discrete energy depositions. The triggered pulses are then processed by the optimum filter technique \cite{Gatti:1986} to evaluate the signal amplitude (i.e. energy) and several pulse-shape parameters. 
In the reconstruction of the coincidences between the LMO and LD, we account for the LD faster response and correct for its constant time shift with respect to the LMO signal, similarly to the method described in \cite{Piperno:2011}. 
%

\section{Results}
\label{sec:results} 

\subsection{Detector performance} 
\label{sec:performance} 

\begin{table}
 \caption{Performance of a scintillating bolometer based on the $\oslash$44 mm Ge light detector coupled to the 45 mm side \LMOenr cubic-shaped scintillator. We report the detectors rise and decay times, the signal amplitude per unit of deposited energy, the energy resolution (FWHM) of the baseline after the optimum filter, at 5.9 keV X-ray of $^{55}$Mn (LD), and at 2615 keV $\gamma$ quanta of $^{208}$Tl  (LMO). We skip the computation of FWHM at 2615 keV for the 18 mK dataset due to poor statistics of the $\gamma$ peak. 
 Particle identification parameters (defined in Sec.~\ref{sec:PID}) as light yield for $\gamma$($\beta$)s $LY_{\gamma(\beta)}$ and a quenching factor for $\alpha$ particles $QF_{\alpha}$, as well as the discrimination power between $\alpha$ and $\gamma$($\beta$) distributions $DP_{\alpha/\gamma(\beta)}$ for events selected in the 2.0--5.1 MeV energy range are also quoted.}
\footnotesize
\begin{center}
\begin{tabular}{llll}
 \hline
Channel & Parameter							& 18 mK & 12 mK \\
 \hline
LD  & Rise time (ms)						& 1.7  & 2.8 \\
~   & Decay time (ms)						& 9.2  & 8.6 \\
~ & ~ & ~ & ~ \\
~   & Signal ($\mu$V/keV)					& 1.20 & 1.44 \\
~ & ~ & ~ & ~ \\
~   & FWHM (keV) at baseline				& 0.300(1) & 0.210(1) \\
~   & FWHM (keV) at 5.9 keV X-ray			& 0.282(5) & 0.315(4) \\
 \hline
LMO & Rise time (ms)			    		& 18 & 25 \\
~   & Decay time (ms)						& 150 & 160 \\
~ & ~ & ~ & ~ \\
~   & Signal ($\mu$V/keV)					& 0.017 & 0.036 \\
~ & ~ & ~ & ~ \\
~   & FWHM (keV) at baseline				& 4.2(2) & 2.5(1) \\
~   & FWHM (keV) at 2615 keV $\gamma$       & --  & 6.0(5)  \\
 \hline
LMO & $LY_{\gamma(\beta)}$ (keV/MeV)	    & 0.635(2) & 0.638(1) \\
\& LD & $QF_{\alpha}$ ($^{210}$Po)		    & 0.192(1) & 0.199(4) \\
~   & $DP_{\alpha/\gamma(\beta)}$			& 7.4(4) & 7.9(1) \\

 \hline
 \end{tabular}
  \label{tab:performance}
 \end{center}
 \end{table}

\normalsize 

\nopagebreak

The performance parameters achieved by the LMO and LD in the 18 and 12 mK tests are listed in Table~\ref{tab:performance}. The rise and decay time constants, defined respectively as time intervals of the (10--90)\% rising edge and (90--30)\% trailing edge relative to the signal maximum, of the LMO are $\sim$0.02 and $\sim$0.15 s, respectively. 
We expect the LD response to be faster by an order of magnitude because of the smaller heat capacity of both the Ge absorber and the NTD thermistor. The time constants obtained are in agreement with the results of previous investigations of similar size LMOs and LDs \cite{Armengaud:2017,Armengaud:2020b}. It is worth noting that the time response of the bolometric detectors depend on the operation temperature and the sensor polarization (see, e.g., \cite{Beeman:2013b}). However, optimization of the detector time response\footnote{In particular, to get the fastest response in view of the rejection of random coincidence events induced background in the $^{100}$Mo $0\nu2\beta$ region of interest \cite{Chernyak:2012,Chernyak:2014,Chernyak:2017}.} was out of the scope of the present study. 

\begin{figure}[htbp]
\begin{center}
\resizebox{0.6\textwidth}{!}{\includegraphics{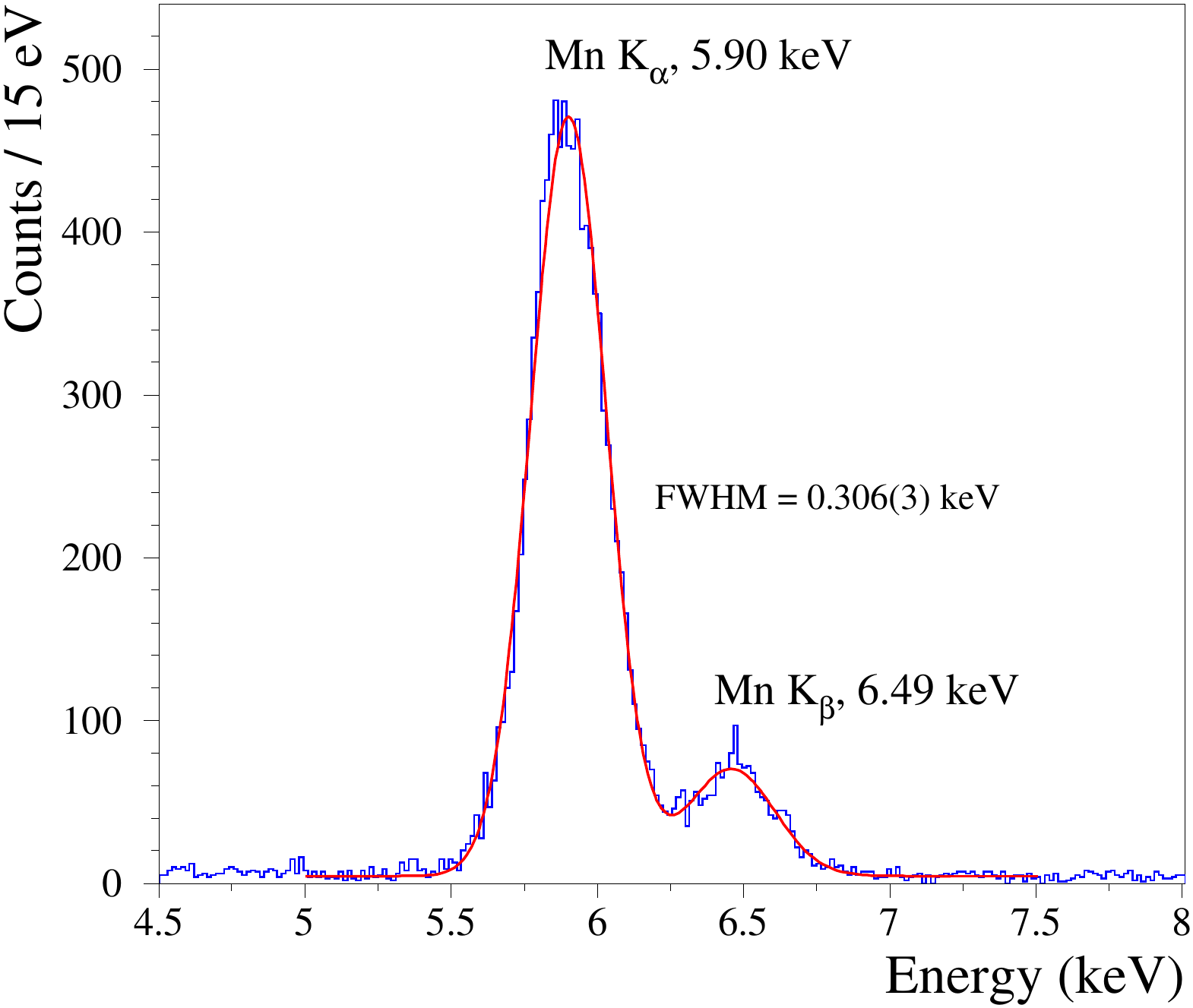}}
\end{center}
\caption{The energy spectrum of the $^{55}$Fe X-ray source measured by a 1.4 g Ge bolometric light  detector over 913 h (12 and 18 mK data) in the CROSS pulse tube based cryogenic facility at the Canfranc underground laboratory (Spain). A fit to the data by a model assuming a double-Gaussian function and a flat background component is shown by solid (red) line. The energy resolution (FWHM) is quoted for the 5.9 keV X-ray peak of Mn K$_\alpha$.}
 \label{fig:LD_55Fe}
\end{figure}

\nopagebreak

The LD signal amplitude per unit of deposited energy is 1.2 $\mu$V/keV at 18 mK and 1.4 $\mu$V/keV at 12 mK. The LMO signal amplitude is of course inferior, 17 nV/keV at 18 mK and doubles at 12 mK. Since the working points were not optimized to get the highest sensitivity, these results are good but not extraordinary among similar devices \cite{Armengaud:2017,Armengaud:2020b}.

The LD is calibrated with the 5.9 and 6.5 keV X-rays emitted by the $^{55}$Fe source. The energy spectrum of the $^{55}$Fe source gathered over 913 h of physics and thorium calibration runs is shown in Fig.~\ref{fig:LD_55Fe}. The almost fully resolved Mn K$_\alpha$/K$_\beta$ doublet is visible thanks to the high LD energy resolution: $\approx$0.3 keV FWHM at 5.9 keV. The baseline noise is 0.2--0.3 keV FWHM, demonstrating a reasonably low threshold.
It is worth noting that such devices do not always show a high energy resolution even if characterized by ten(s) eV RMS noise\footnote{For example, the Mn doublet resolution of 0.3--0.5 keV FWHM was measured with LDs made of 30--45 $\mu$m thick Ge wafers \cite{Armengaud:2017}, while the 0.08 keV FWHM resolution was achieved by a 33 g Ge bolometer ($\oslash$20$\times$20 mm) characterized by a similar noise level \cite{Armengaud:2019}.}, due to the position-dependent response of thin bolometers.

\begin{figure}[htbp]
\begin{center}
\resizebox{0.7\textwidth}{!}{\includegraphics{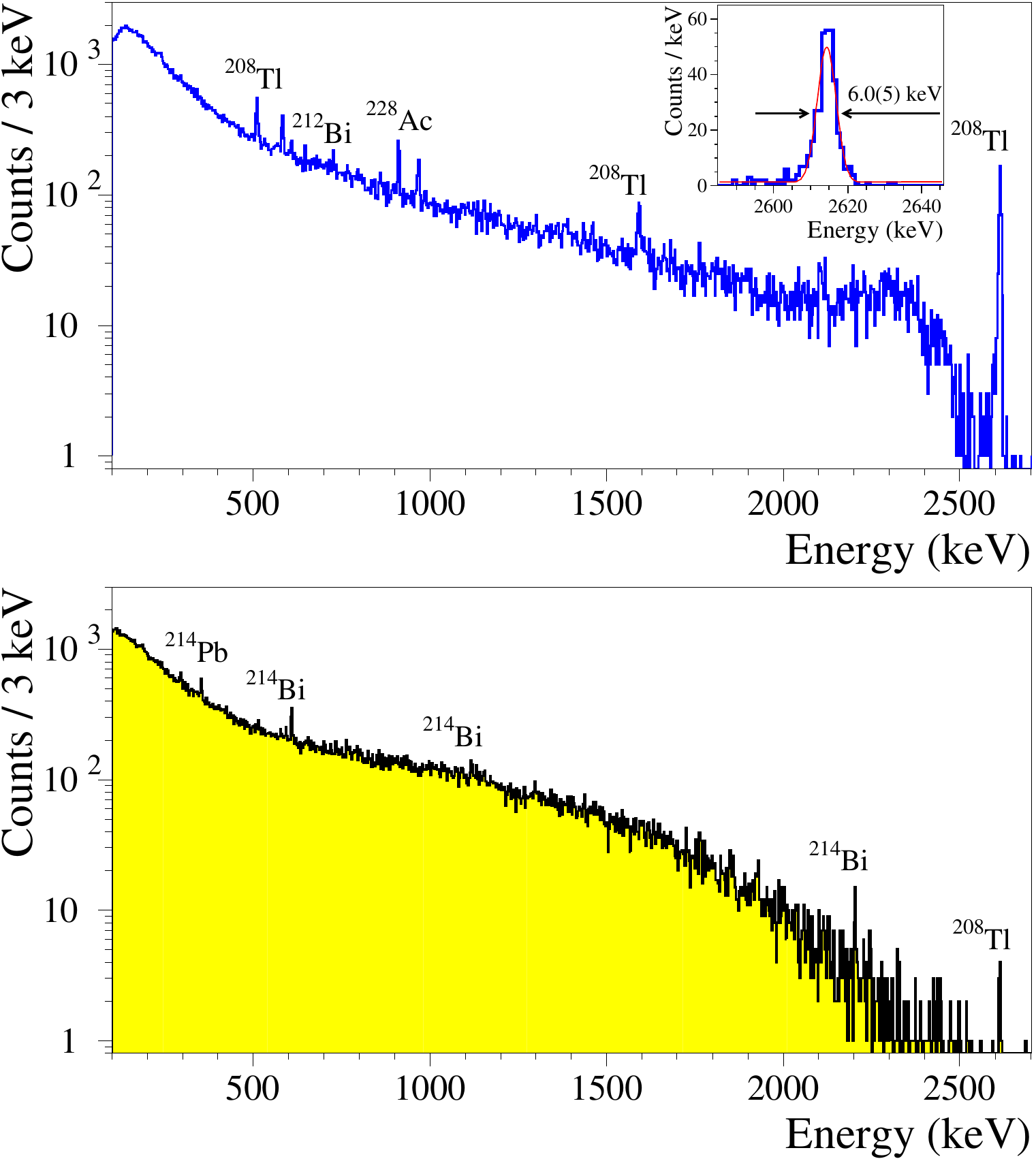}}
\end{center}
\caption{The energy spectra of $\gamma$($\beta$) events accumulated by the LMO over the calibration (285 h; top panel) and physics data (628 h; bottom panel) measurements in the C2U facility at the Canfranc underground laboratory. The most prominent $\gamma$ peaks are labeled. The contribution of $\alpha$ events to the physics data has been removed with the scintillation light based particle identification (see text). The inset shows the $\gamma$ peak with energy of 2615 keV in the calibration data together with a fit and the calculated energy resolution (FWHM).}
 \label{fig:LMO_spectra}
\end{figure} 

\nopagebreak

We measured the LMO energy scale and resolution with the most intense gamma peaks in the thorium spectrum, illustrated in Fig. \ref{fig:LMO_spectra} (top panel). Because of the incomplete shielding, the physics data (Fig. \ref{fig:LMO_spectra}, bottom panel) also exhibit several $\gamma$ peaks from residual environmental radioactivity (daughters of $^{226}$Ra and $^{228}$Th sub-chains).  The $^{238}$U/$^{234}$U alpha source also emits $\beta$ particles from $^{234m}$Pa decays ($Q_{\beta}$ = 2.27 MeV \cite{Audi:2017}), which, together with the $^{100}$Mo $2\nu2\beta$ decays \cite{Armengaud:2020b}, are responsible for the most part of the continuum background above 0.5 MeV, seen in Fig. \ref{fig:LMO_spectra} (bottom panel). 

\nopagebreak

\begin{figure}[htbp]
\begin{center}
\resizebox{0.6\textwidth}{!}{\includegraphics{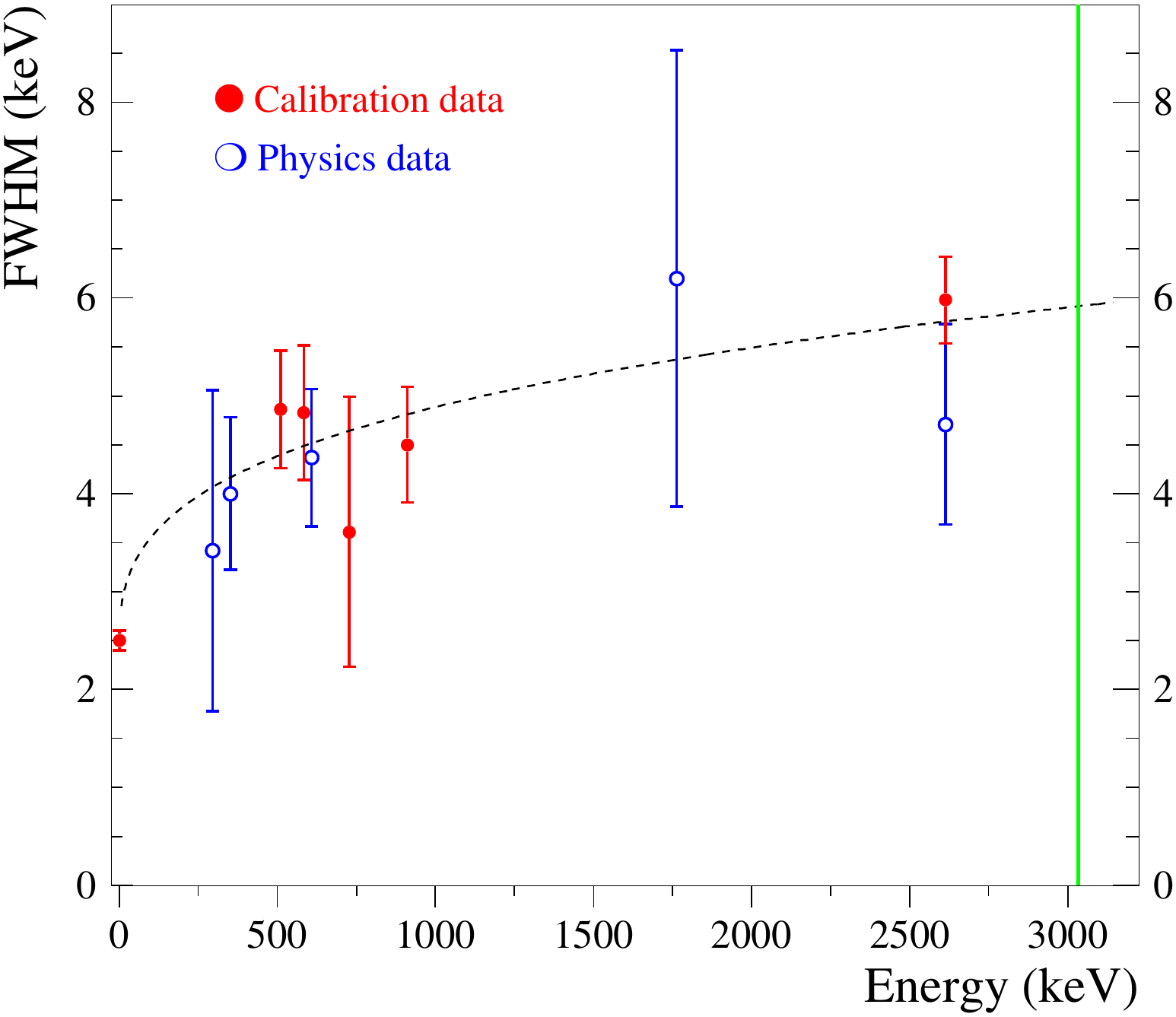}}
\end{center}
\caption{Energy dependence of the 279 g \LMOenr bolometer energy resolution (FWHM) measured in the  calibration (filled circles; 12 mK) and physics (open circles; 18 and 12 mK) runs. The fitting curve is shown by the dashed line, while the solid line indicates the $^{100}$Mo $Q_{2\beta}$.}
 \label{fig:LMO_FWHM}
\end{figure}

The energy dependence of the LMO energy resolution is presented in Fig. \ref{fig:LMO_FWHM}. The results extracted from physics data are limited by the poor statistics. The detector demonstrates a good energy resolution in a wide energy interval exhibiting a peak width slightly increasing with energy, in agreement with early findings \cite{Armengaud:2017,Schmidt:2020}. 
In particular, we achieved a 6 keV energy resolution (FWHM) for $\gamma$-ray quanta of $^{208}$Tl with energy 2615 keV, and a 2.5 keV FWHM baseline noise. The resolution at the $Q_{2\beta}$ of $^{100}$Mo is expected to be very similar (Fig. \ref{fig:LMO_FWHM}). These results are in agreement with prior measurements for cylindrical LMOs  \cite{Armengaud:2017,Poda:2017a,Armengaud:2020a}, confirming an excellent bolometric performance independent of the crystal shape. It is also evident that a low baseline noise is crucial in obtaining a high energy resolution with a \LMOenr bolometer.  It is worth noting, the lowest noise achieved with the LMO is a factor 2--4 worse than the best reported values for large-volume lithium molybdate bolometers \cite{Armengaud:2017,Armengaud:2020a}. Thus, taking into account sub-optimal noise level of the present study, there is still room for improvement.

\nopagebreak

\subsection{Particle identification capability} 
\label{sec:PID} 

\begin{figure}[htbp]
\begin{center}
\resizebox{0.6\textwidth}{!}{\includegraphics{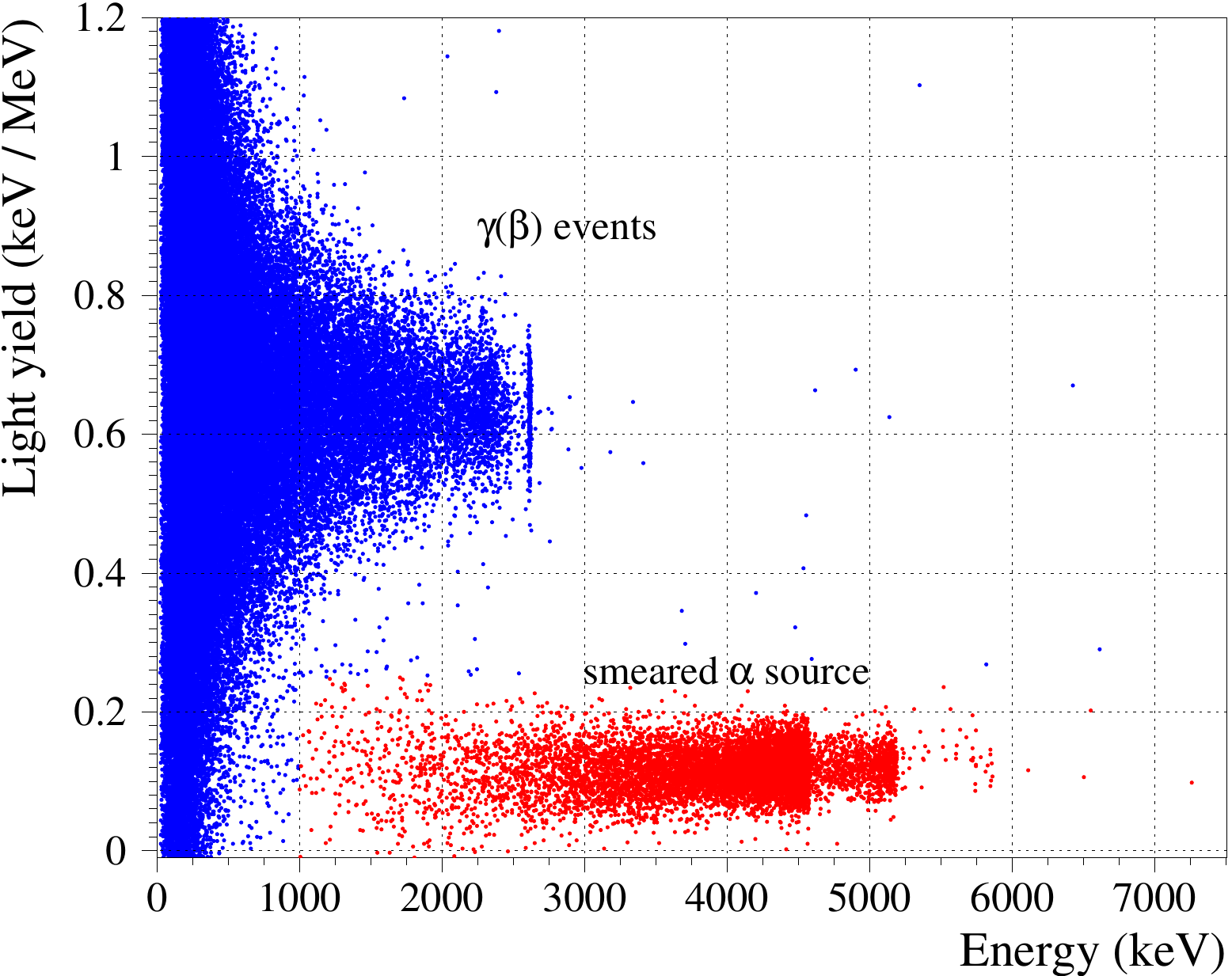}}
\end{center}
\caption{Light yield as a function of energy deposited in the LMO and measured in a 125 h long $^{232}$Th calibration at 12 mK. The energy scale is calibrated with $\gamma$ quanta. The population of $\gamma$($\beta$) events is clearly separated from the $\alpha$ events, mainly originated by the $^{238}$U/$^{234}$U smeared $\alpha$ source. The $\alpha$ events shown in red were selected above 1 MeV with a $LY$ cut below 0.25 keV/MeV.}
 \label{fig:LMO_Qplot}
\end{figure}

Coincidences between the LMO and LD have been used to probe the scintillation based particle identification (PID). For each event triggered by the LMO we calculated a PID parameter, the so-called light yield ($LY$), defined as the ratio of the LD to LMO measured energy.  
The $LY$ dependence on particle energy is shown in Fig.~\ref{fig:LMO_Qplot}, where the population of $\gamma$($\beta$) events is clearly separated from $\alpha$'s. Such a powerful separation is achieved thanks to the quenching of the scintillation light for $\alpha$ particles with respect to $\gamma$($\beta$)'s of the same energy and low noise of the LD. Different ionization properties lead also to a different amplitude measured by the LMO. Figure~\ref{fig:LMO_Qplot} shows a $\sim$7\% increase of an $\alpha$ event energy with respect to the gamma energy scale, in agreement with previous studies of LMO bolometers \cite{Armengaud:2017,Armengaud:2020a}. This difference, called thermal quenching, hints at a possibility of PID by pulse-shape analysis of the heat channel itself \cite{Armengaud:2017}, but it is by far less reproducible due to a strong dependence on the noise conditions\footnote{Since the present detector does not have the CROSS technology of the surface coating for PID purpose, we skip an analysis of pulse-shape discrimination of $\alpha$ events.}.

\nopagebreak

\begin{figure}[htbp]
\begin{center}
\resizebox{0.6\textwidth}{!}{\includegraphics{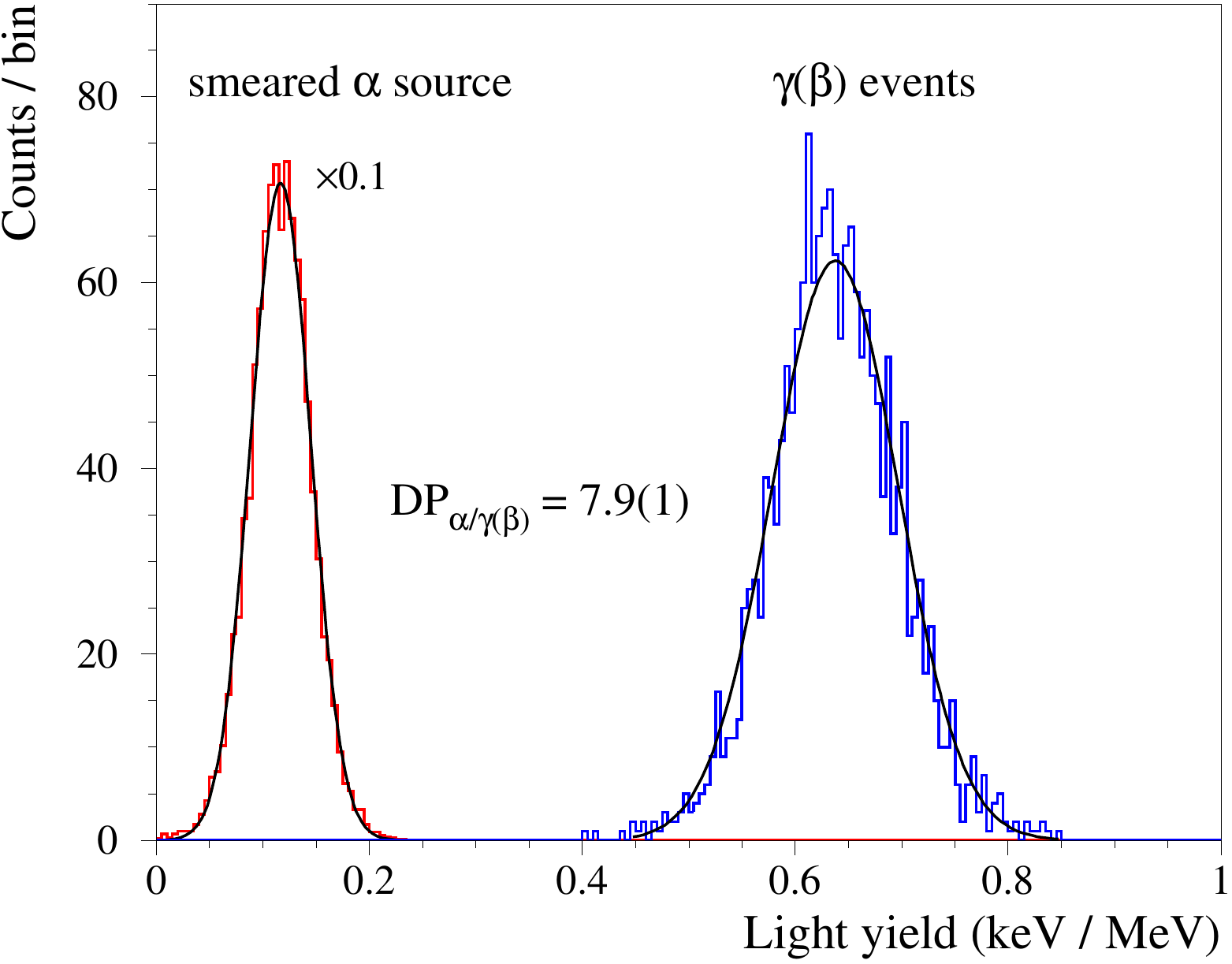}}
\end{center}
\caption{The light yield distributions for $\alpha$ and $\gamma$($\beta$) particles, selected in the 2.0--5.1 MeV energy interval. The former distribution is rescaled by a factor 0.1 to improve the visibility of the later one. A Gaussian fit to each distribution is shown by a solid line. The discrimination power between the two populations is $\sim$8.}
 \label{fig:LMO_DP}
\end{figure} 

\nopagebreak

In order to investigate a $LY$-based $\alpha$/$\gamma$($\beta$) separation close to the $^{100}$Mo $0\nu2\beta$ region of interest (ROI), we selected events within the 2.0--5.1 MeV energy range and a $LY$ interval of 0.4--1.0 keV/MeV for $\gamma$($\beta$)'s and 0--0.25 keV/MeV for $\alpha$'s; both are shown in Fig.~\ref{fig:LMO_DP}. A Gaussian fit to each distribution provides a $LY$ mean value ($\mu$) and standard deviation ($\sigma$), which we used to calculate the so-called discrimination power defined as 
$DP_{\alpha/\gamma(\beta)} = \left| \mu_{\gamma(\beta)} - \mu_{\alpha} \right| / \sqrt{\sigma_{\gamma(\beta)}^2 + \sigma_{\alpha}^2}$. The $DP_{\alpha/\gamma(\beta)}$ based on the present data is around 8 (Table~\ref{tab:performance}), meaning about 8$\sigma_{\alpha}$ of alpha event rejection while keeping almost 100\% of $\gamma$($\beta$)'s. This rejection power fully satisfies the CUPID goal of identifying 99.9\% of alpha particles (corresponding to $DP_{\alpha/\gamma(\beta)}$ $\sim$ 3.1).

The LMO light yield for $\gamma$($\beta$) events ($LY_{\gamma(\beta)}$) was found to be 0.64 keV/MeV, similar to one measured with cylindrical shaped LMOs of $\oslash$44$\times$45 mm size  \cite{Armengaud:2017,Armengaud:2020a,Poda:2020}. However, in the present study the light collection efficiency was affected by the smaller area of the  LD (15 cm$^2$) with respect to the LMO surface facing it (20 cm$^2$) and by the entrance window of the Cu holder ($\oslash$45 mm). Thus, considering only the direct light, the $LY_{\gamma(\beta)}$ is expected to be $\sim$ 0.85 keV/MeV, once a 45~mm side square LD is coupled to the LMO. In case of the use of two LDs, the $LY$ should be roughly doubled, as demonstrated by CUPID-Mo \cite{Armengaud:2020a,Poda:2020}. The quenching factor ($QF_{\alpha}$) for $\alpha$ particles of $^{210}$Po observed in the data (see the next section) is 0.2~\footnote{Such a parameter is typically quoted without the correction from the thermal quenching and we follow that convention here.}, in agreement with the previous data \cite{Armengaud:2017,Armengaud:2020a,Poda:2020}.

\subsection{\LMOenr crystal radiopurity}
\label{sec:radiopurity} 

\begin{figure}[htbp]
\begin{center}
\resizebox{0.6\textwidth}{!}{\includegraphics{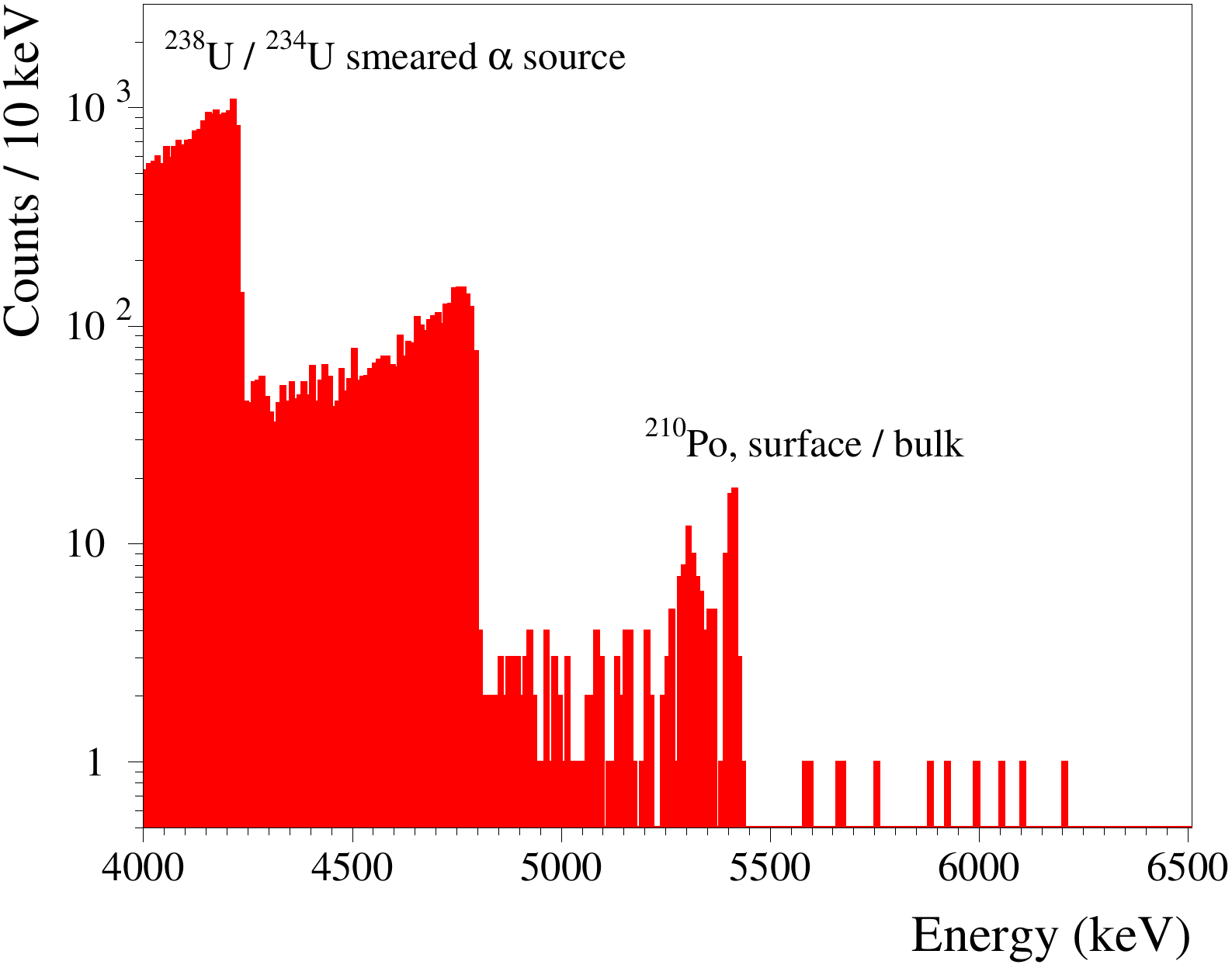}}
\end{center}
\caption{A part of the $\alpha$ energy spectrum accumulated by the LMO operated over 628 h in the CROSS underground cryostat in Canfranc. The energy interval covers the $Q_{\alpha}$-values of the most $\alpha$-active radionuclides from U/Th decay chains. In addition to a dominant contribution from the used $^{238}$U/$^{234}$U source, the spectrum exhibits only two populations of $^{210}$Po originated by external (surface of nearby materials) or internal (crystal bulk) contaminations of the detector.}
 \label{fig:LMO_Bkg_alpha}
\end{figure}

A highly-efficient PID together with a good energy resolution of the LMO operated over four weeks of background measurements allow us to quantify the Li$_{2}${}$^{100}$MoO$_4$ radiopurity with a high sensitivity. The spectrum of alpha events selected from the physics data (and recalibrated to $\alpha$ energy) is presented in Fig. \ref{fig:LMO_Bkg_alpha}; the energy interval covers most $Q_{\alpha}$-values of radionuclides from the U/Th chains. As it is seen in Fig.  \ref{fig:LMO_Bkg_alpha}, the use of the $^{238}$U/$^{234}$U smeared $\alpha$ source prevents estimation of the alpha activity of $^{232}$Th, $^{238}$U and some of the daughters with $Q_{\alpha}$ $\leq$ 4.8 MeV. However, we can investigate a possible contamination by $^{226}$Ra and $^{228}$Th, which are the most harmful contaminants for $0\nu2\beta$  searches.

The energy region above 4.8 MeV contains only two peak-like structures both ascribed to $^{210}$Po $\alpha$ events, and originated by the $^{210}$Pb contamination \cite{Armengaud:2015}. A clear peak at 5.4 MeV is induced by the $^{210}$Po decays in the \LMOenr crystal bulk with the activity of 80(12) $\mu$Bq/kg. A $^{210}$Po ($^{210}$Pb) bulk contamination on the level of ten(s)--hundred(s) $\mu$Bq/kg is typical for \LMOenr crystals \cite{Armengaud:2017,Poda:2017a,Armengaud:2020a,Poda:2020}.
A broad distribution peaked at 5.3 MeV is caused by the $^{210}$Po decay on the surface of materials facing the LMO (a 0.1 MeV energy is taken away by the $^{206}$Pb nuclear recoil). The decays of $^{210}$Po at surfaces of the detector materials can populate the $^{100}$Mo $0\nu2\beta$ ROI as energy-degraded alpha events, but they can be easily rejected thanks to the efficient PID of \LMOenr scintillating bolometers. Also, the total rate of $^{210}$Po events ($\sim$0.06 mHz) is rather low to be a notable source of pile-ups, which are of certain concern for slow response thermal detectors \cite{Poda:2017}. 

\nopagebreak

The spectrum shows no other structures, so we can only set limits on other radionuclide contaminations. In order to be conservative, we considered all events within 25 keV of each expected peak location. The background (1 count per 50 keV) was estimated in the 5.65--5.75 and 5.85--5.95 MeV energy intervals, containing no $Q_{\alpha}$-value of U/Th radionuclides. The data exhibit no events of $^{228}$Th ($Q_{\alpha}$ = 5520 keV \cite{Wang:2017a}) and two counts of $^{222}$Rn ($Q_{\alpha}$ = 5590 keV \cite{Wang:2017a}; a daughter of $^{226}$Ra), thus we place 90 confidence level (C.L.) upper limits of 1.6 and 4.9 counts, respectively \cite{Feldman:1998}. 
A selection efficiency of 95.7\% was estimated using alpha events of the $^{238}$U/$^{234}$U smeared $\alpha$ source, distributed in the 3.5--4.8 MeV energy interval. Therefore, the activity of $^{228}$Th and $^{226}$Ra in the \LMOenr crystal bulk is below 3 and 8 $\mu$Bq/kg at 90\% C.L., respectively. Taking into account the reasonably short half-lives of $^{228}$Ra (5.75 yr \cite{Audi:2017}) and $^{228}$Th (1.91 yr \cite{Audi:2017}), the limit on the $^{228}$Th activity can also represent the $^{232}$Th contamination in the crystals, as e.g. seen in \cite{Armengaud:2017,Schmidt:2020,Poda:2020}. 

The limits on the $^{228}$Th and $^{226}$Ra activity in the studied \LMOenr crystal are on the same level as reported by LUMINEU \cite{Armengaud:2017,Poda:2017a}, which were obtained by the analysis of a comparable exposure. A significantly larger exposure of the CUPID-Mo experiment \cite{Armengaud:2020a,Schmidt:2020,Poda:2020} shows that 
the level of remaining contaminants in \LMOenr scintillators can be even an order of magnitude lower. As the protocol of the CROSS crystal production was the one adopted by LUMINEU and CUPID-Mo, it seems natural to assume the radiopurity of the CROSS crystals to be similar to that of CUPID-Mo ones. It is worth noting that the $^{228}$Th ($^{232}$Th) and $^{226}$Ra contamination on the level of 10~$\mu$Bq/kg is compatible with a background contribution below 10$^{-4}$ counts/yr/kg/keV to the $^{100}$Mo $0\nu2\beta$ ROI, and it is fully acceptable not only for the medium-scale CROSS experiment (with $\sim$10~kg detector mass), but also for the tonne-scale extension of CUPID.

\section{Conclusions}

We report that performance and radiopurity of a scintillating bolometer based on a large cubic-shaped Li$_{2}${}$^{100}$MoO$_4$ crystal ---randomly taken from 32 identical crystals (with a 45 mm side and a mass of 0.28 kg each) of the CROSS \onbb project---  are similar to those of cylindrical 0.2~kg Li$_{2}${}$^{100}$MoO$_4$ bolometers, used in LUMINEU and  CUPID-Mo $2\beta$ experiments. 
In particular, the \LMOenr detector energy resolution at 2615 keV $\gamma$ quanta (as well as its approximation to 3034 keV, the $^{100}$Mo $2\beta$ decay energy) is 6 keV FWHM. A scintillation light yield of the cubic-shaped crystal, 0.64~keV/MeV for $\gamma$($\beta$)'s, is compatible to that of cylindrical crystals. However, the measured light yield is affected by about 30\% lower detection surface of the optical bolometer with respect to the nearby \LMOenr crystal face.
In spite of sub-optimal light collection, a full separation ($\sim$8$\sigma$) between $\alpha$ and $\gamma$($\beta$) events above 2 MeV has been achieved. A high radiopurity of the cubic-shaped \LMOenr crystal was also demonstrated by the present study, where only $^{210}$Po is detected with the activity of 80(12) $\mu$Bq/kg, while the content of $^{228}$Th and $^{226}$Ra (the most harmful radionuclides from U/Th families for \onbb searches) is estimated to be less than 3 and 8~$\mu$Bq/kg, respectively.

The performed investigation additionally proves the excellent prospects of Li$_{2}${}$^{100}$MoO$_4$  scintillating bolometers for high-sensitivity \onbb decay  searches. The cubic shape of large-volume ($\sim$90 cm$^3$) \LMOenr crystals allows a more compact detector array structure and thus the deployment of a larger isotope mass in the experimental volume, as well as an increased efficiency of multi-site event detection. In view of these results, large-mass ($\sim$0.3 kg) radiopure cubic-shaped \LMOenr crystals operated as bolometers satisfy the demands of the CROSS and CUPID projects.

\acknowledgments

The CROSS and CUPID Collaborations thank the directors and staff of the Laboratorio Subterr\'aneo de Canfranc and the technical staff of our laboratories. This work was supported by the Istituto Nazionale di Fisica Nucleare (INFN); by the European Research Council (ERC) under the European Union Horizon 2020 program (H2020/2014-2020) with the ERC Advanced Grant no. 742345 (ERC-2016-ADG, project CROSS) and the Marie Sklodowska-Curie Grant Agreement No. 754496; by the Italian Ministry
of University and Research (MIUR) through the grant Progetti di ricerca di Rilevante Interesse Nazionale (PRIN 2017, grant no. 2017FJZMCJ); by the US National Science Foundation under Grant
Nos. NSF-PHY-1614611 and NSF-PHY-1401832;  
by the P2IO LabEx (ANR-10-LABX-0038) managed by the Agence Nationale de la Recherche (France). 
This material is also based upon work supported by the US Department of Energy (DOE) Office of
Science under Contract Nos. DE-AC02-05CH11231 and DE-AC02-06CH11357; and by the DOE Office of Science, Office of Nuclear Physics under Contract Nos. DE-FG02-08ER41551, DE-SC0011091, DE-SC0012654, DE-SC0019316, DE-SC0019368, and DE-SC0020423. This work was also supported by the Russian Science Foundation under grant No. 18-12-00003 and the National Research Foundation of Ukraine under Grant No. 2020.02/0011.


\providecommand{\href}[2]{#2}\begingroup\raggedright\endgroup



\begin{thebibliography}{10}

\bibitem{Vergados:2016}
J.~D. Vergados, H.~Ejiri and F.~Simkovic, \emph{{Neutrinoless double beta decay
  and neutrino mass}}, {\emph{Int. J. Mod. Phys. E} {\bfseries 25} (2016)
  1630007}.

\bibitem{Dolinski:2019}
M.~J. Dolinski, A.~W.~P. Poon and W.~Rodejohann, \emph{{Neutrinoless
  Double-Beta Decay: Status and Prospects}}, {\emph{Annu. Rev. Nucl. Part.
  Sci.} {\bfseries 69} (2019) 219}.

\bibitem{Barabash:2020}
A.~S. Barabash, \emph{{Precise Half-Life Values for Two-Neutrino Double-$\beta$
  Decay: 2020 Review}}, {\emph{Universe} {\bfseries 6} (2020) 159}.

\bibitem{Pirro:2017}
S.~Pirro and P.~Mauskopf, \emph{{Advances in Bolometer Technology for
  Fundamental Physics}}, {\emph{Annu. Rev. Nucl. Part. Sci.} {\bfseries 67}
  (2017) 161}.

\bibitem{Poda:2017a}
{\scshape {LUMINEU}, {EDELWEISS}, and {CUPID-0/Mo}} collaboration,
  \emph{{$^{100}$Mo-enriched Li$_2$MoO$_4$ scintillating bolometers for $0\nu
  2\beta$ decay search: from LUMINEU to CUPID-0/Mo projects}}, {\emph{AIP Conf.
  Proc.} {\bfseries 1894} (2017) 020017}.

\bibitem{Bellini:2018}
F.~Bellini, \emph{Potentialities of the future technical improvements in the
  search of rare nuclear decays by bolometers}, {\emph{Int. J. Mod. Phys. A}
  {\bfseries 33} (2018) 1843003}.

\bibitem{Adams:2020}
{\scshape {CUORE}} collaboration, \emph{{Improved Limit on Neutrinoless
  Double-Beta Decay in $^{130}$Te with CUORE}}, {\emph{Phys. Rev. Lett.}
  {\bfseries 124} (2020) 122501}.

\bibitem{Wang:2017a}
M.~Wang et~al., \emph{{The AME2016 atomic mass evaluation $^{\ast}$ (II).
  Tables, graphs and references}}, {\emph{Chinese Phys. C} {\bfseries 41}
  (2017) 030003}.

\bibitem{Meija:2016}
J.~Meija et~al., \emph{{Isotopic compositions of the elements 2013 (IUPAC
  Technical Report)}}, {\emph{Pure Appl. Chem.} {\bfseries 88} (2016) 293}.

\bibitem{Bellini:2010}
F.~Bellini et~al., \emph{{Response of a TeO$_2$ bolometer to $\alpha$
  particles}}, {\emph{JINST} {\bfseries 5} (2010) P12005}.

\bibitem{Wang:2015raa}
{\scshape CUPID} collaboration, \emph{{CUPID: CUORE (Cryogenic Underground
  Observatory for Rare Events) Upgrade with Particle IDentification}},
  {\emph{\href{https://arxiv.org/abs/1504.03599}} (2015) }
  [\href{https://arxiv.org/abs/1504.03599}{{\ttfamily 1504.03599}}].

\bibitem{Chernyak:2012}
D.~M. Chernyak et~al., \emph{{Random coincidence of $2 \nu 2 \beta$ decay
  events as a background source in bolometric $0 \nu 2 \beta$ decay
  experiments}}, {\emph{Eur. Phys. J. C} {\bfseries 72} (2012) 1989}.

\bibitem{Chernyak:2014}
D.~M. Chernyak et~al., \emph{{Rejection of randomly coinciding events in
  ZnMoO$_4$ scintillating bolometers}}, {\emph{Eur. Phys. J. C} {\bfseries 74}
  (2014) 2913}.

\bibitem{Chernyak:2017}
D.~M. Chernyak et~al., \emph{{Rejection of randomly coinciding events in
  Li$_2${}$^{100}$MoO$_4$ scintillating bolometers using light detectors based
  on the Neganov-Luke effect}}, {\emph{Eur. Phys. J. C} {\bfseries 77} (2017)
  3}.

\bibitem{Wang:2015taa}
{\scshape CUPID} collaboration, \emph{{R\&D towards CUPID (CUORE Upgrade with
  Particle IDentification)}}, {\emph{\href{https://arxiv.org/abs/1504.03612}}
  (2015) } [\href{https://arxiv.org/abs/1504.03612}{{\ttfamily 1504.03612}}].

\bibitem{Poda:2017}
D.~Poda and A.~Giuliani, \emph{Low background techniques in bolometers for
  double-beta decay search}, {\emph{Int. J. Mod. Phys. A} {\bfseries 32} (2017)
  1743012}.

\bibitem{Azzolini:2019tta}
{\scshape CUPID-0} collaboration, \emph{{Final Result of CUPID-0 Phase-I in the
  Search for the $^{82}$Se Neutrinoless Double-$\beta$ Decay}}, {\emph{Phys.
  Rev. Lett.} {\bfseries 123} (2019) 032501}.

\bibitem{Armengaud:2017}
E.~Armengaud et~al., \emph{{Development of $^{100}$Mo-containing scintillating
  bolometers for a high-sensitivity neutrinoless double-beta decay search}},
  {\emph{Eur. Phys. J. C} {\bfseries 77} (2017) 785}.

\bibitem{Helis:2020}
D.~L. Helis et~al., \emph{{Neutrinoless double-beta decay searches with
  enriched $^{116}$CdWO$_4$ scintillating bolometers}}, {\emph{J. Low Temp.
  Phys.} {\bfseries 199} (2020) 467}.

\bibitem{Giuliani:2018}
A.~Giuliani, F.~A. Danevich and V.~I. Tretyak, \emph{{A multi-isotope $0\nu
  2\beta $ bolometric experiment}}, {\emph{Eur. Phys. J. C} {\bfseries 78}
  (2018) 272}.

\bibitem{CUPIDInterestGroup:2019inu}
{\scshape CUPID} collaboration, \emph{{CUPID pre-CDR}},
  {\emph{\href{https://arxiv.org/abs/1907.09376}} (2019) }
  [\href{https://arxiv.org/abs/1907.09376}{{\ttfamily 1907.09376}}].

\bibitem{Grigorieva:2017}
V.~D. Grigorieva et~al., \emph{{Li$_2$MoO$_4$ crystals grown by low thermal
  gradient Czochralski technique}}, {\emph{J. Mat. Sci. Eng. B} {\bfseries 7}
  (2017) 63}.

\bibitem{Armengaud:2020b}
E.~Armengaud et~al., \emph{{Precise measurement of $2\nu\beta\beta$ decay of
  $^{100}$Mo with the CUPID-Mo detection technology}}, {\emph{Eur. Phys. J. C}
  {\bfseries 80} (2020) 674}.

\bibitem{Armengaud:2020a}
E.~Armengaud et~al., \emph{{The CUPID-Mo experiment for neutrinoless
  double-beta decay: performance and prospects}}, {\emph{Eur. Phys. J. C}
  {\bfseries 80} (2020) 44}.

\bibitem{Schmidt:2020}
{\scshape {CUPID-Mo}} collaboration, \emph{{First data from the CUPID-Mo
  neutrinoless double beta decay experiment}}, {\emph{J. Phys.: Conf. Ser.}
  {\bfseries 1468} (2020) 012129}.

\bibitem{Poda:2020}
{\scshape {CUPID-Mo}} collaboration, \emph{{Performance of the CUPID-Mo
  double-beta decay bolometric experiment}},  in \emph{{XXIX International
  (online) Conference on Neutrino Physics and Astrophysics (Neutrino 2020),
  June 22 -- July 02, 2020}}, 2020,
  \href{https://indico.fnal.gov/event/19348/contributions/186385/}{https://indico.fnal.gov/event/19348/contributions/186385/}.

\bibitem{Bandac:2020}
{\scshape CROSS} collaboration, \emph{{The 0$\nu$2$\beta$-decay CROSS
  experiment: preliminary results and prospects}}, {\emph{JHEP} {\bfseries 01}
  (2020) 018}.

\bibitem{Khalife:2020}
H.~Khalife et~al., \emph{{The CROSS Experiment: Rejecting Surface Events by PSD
  Induced by Superconducting Films}}, {\emph{J. Low Temp. Phys.} {\bfseries
  199} (2020) 19}.

\bibitem{Khalife:2020a}
{\scshape {CROSS}} collaboration, \emph{{The CROSS experiment: rejecting
  surface events with PSD}},  in \emph{{XXIX International (online) Conference
  on Neutrino Physics and Astrophysics (Neutrino 2020), June 22 -- July 02,
  2020}}, 2020,
  \href{https://indico.fnal.gov/event/19348/contributions/186492/}{https://indico.fnal.gov/event/19348/contributions/186492/}.

\bibitem{Zolotarova:2020a}
{\scshape {CROSS}} collaboration, \emph{{First results of CROSS underground
  measurements with massive bolometers}},  in \emph{{XXIX International
  (online) Conference on Neutrino Physics and Astrophysics (Neutrino 2020),
  June 22 -- July 02, 2020}}, 2020,
  \href{https://indico.fnal.gov/event/19348/contributions/186216/}{https://indico.fnal.gov/event/19348/contributions/186216/}.

\bibitem{Haller:1994}
E.~E. Haller, \emph{{Advanced far-infrared detectors}}, {\emph{Infrared Phys.
  Techn.} {\bfseries 35} (1994) 127}.

\bibitem{Andreotti:2012}
E.~Andreotti et~al., \emph{{Production, characterization and selection of the
  heating elements for the response stabilization of the CUORE bolometers}},
  {\emph{Nucl. Instrum. Meth. A} {\bfseries 664} (2012) 161}.

\bibitem{Alessandrello:1998}
A.~Alessandrello et~al., \emph{{Methods for response stabilization in
  bolometers for rare decays}}, {\emph{J. Cryst. Growth} {\bfseries 412} (1998)
  454}.

\bibitem{Novati:2018}
V.~Novati, \emph{{Sensitivity enhancement of the CUORE experiment via the
  development of Cherenkov hybrid TeO$_2$ bolometers}}, Ph.D. thesis,
  University Paris-Saclay, Orsay, France, November, 2018.

\bibitem{Berge:2018}
L.~Berg\'e et~al., \emph{{Complete event-by-event $\alpha$/$\gamma$($\beta$)
  separation in a full-size TeO$_2$ CUORE bolometer by Neganov-Luke-magnified
  light detection}}, {\emph{Phys. Rev. C} {\bfseries 97} (2018) 032501}.

\bibitem{Zolotarova:2020}
{\scshape CROSS} collaboration, \emph{The {CROSS} experiment: search for
  0$\nu$2$\beta$ decay with surface sensitive bolometers}, {\emph{J. Phys.:
  Conf. Ser.} {\bfseries 1468} (2020) 012147}.

\bibitem{Olivieri:2020}
{\scshape {CROSS}} collaboration, \emph{{The new CROSS Cryogenic Underground
  (C2U) facility: an overview}},  in \emph{{XXIX International (online)
  Conference on Neutrino Physics and Astrophysics (Neutrino 2020), June 22 --
  July 02, 2020}}, 2020,
  \href{https://indico.fnal.gov/event/19348/contributions/186315/}{https://indico.fnal.gov/event/19348/contributions/186315/}.

\bibitem{Cebrian:2004}
S.~Cebri\'an et~al., \emph{{Bolometric WIMP search at Canfranc with different
  absorbers}}, {\emph{Astropart. Phys.} {\bfseries 21} (2004) 23}.

\bibitem{Olivieri:2017}
E.~Olivieri et~al., \emph{{Vibrations on pulse tube based Dry Dilution
  Refrigerators for low noise measurements}}, {\emph{Nucl. Instrum. Meth. A}
  {\bfseries 858} (2017) 73}.

\bibitem{Arnaboldi:2002}
C.~{Arnaboldi} et~al., \emph{The programmable front-end system for {CUORICINO},
  an array of large-mass bolometers}, {\emph{IEEE Trans. Nucl. Sci.} {\bfseries
  49} (2002) 2440}.

\bibitem{Carniti:2020}
P.~Carniti, C.~Gotti and G.~Pessina, \emph{{High-Resolution Digitization System
  for the CROSS Experiment}}, {\emph{J. Low Temp. Phys.} {\bfseries 199} (2020)
  833}.

\bibitem{Gatti:1986}
E.~Gatti and P.~Manfredi, \emph{Processing the signals from solid-state
  detectors in elementary-particle physics}, {\emph{Riv. Nuovo Cim.} {\bfseries
  9} (1986) 1}.

\bibitem{Piperno:2011}
G.~Piperno, S.~Pirro and M.~Vignati, \emph{{Optimizing the energy threshold of
  light detectors coupled to luminescent bolometers}}, {\emph{JINST} {\bfseries
  6} (2011) P10005}.

\bibitem{Beeman:2013b}
J.~W. Beeman et~al., \emph{{Characterization of bolometric Light Detectors for
  rare event searches}}, {\emph{JINST} {\bfseries 8} (2013) P07021}.

\bibitem{Armengaud:2019}
{\scshape {EDELWEISS}} collaboration, \emph{{Searching for low-mass dark matter
  particles with a massive Ge bolometer operated above ground}}, {\emph{Phys.
  Rev. D} {\bfseries 99} (2019) 082003}.

\bibitem{Audi:2017}
G.~Audi et~al., \emph{{The NUBASE2016 evaluation of nuclear properties}},
  {\emph{Chinese Phys. C} {\bfseries 41} (2017) 030001}.

\bibitem{Armengaud:2015}
E.~Armengaud et~al., \emph{{Development and underground test of radiopure
  ZnMoO$_4$ scintillating bolometers for the LUMINEU $0\nu2\beta$ project}},
  {\emph{JINST} {\bfseries 10} (2015) P05007}.

\bibitem{Feldman:1998}
G.~J. Feldman and R.~D. Cousins, \emph{{Unified approach to the classical
  statistical analysis of small signals}}, {\emph{Phys. Rev. D} {\bfseries
  3873} (1998) 57}.

\end{thebibliography}
\end{document}